\documentclass[11pt]{article}
\usepackage{amsfonts,latexsym, amssymb, amsmath} % symboles
\usepackage{graphicx} % graphiques
\newcounter{cst}

\newcounter{cexp}

\renewenvironment{abstract}{\begin{center} {\bf Abstract} \\
\end{center} }{\smallskip}

\begin{document}

\def\dsp{\displaystyle}

\newcommand\mc{\spadesuit}
\newcommand\gk{\clubsuit}

% <<<<<<< tableaux avec et sans numerotation
\def\be{\begin{equation}}
\def\ee{\end{equation}}

\def\bea{\begin{eqnarray}} %{lllll}}

\def\eea{\end{eqnarray}}

\def\beqsys {\be\ba \left \{ \begin{array}{l}}
\def\eeqsys {\end{array} \right . \ea\ee }

\def\beqsysno {\be\ba \left \{ \begin{array}{l}}
\def\eeqsysno {\end{array} \right . \ea\ee}

%   <<<<<<<< definitions variables

\def\app{{\DD}}
\def\appm{{\DD_m}}

\def\BB{{\cal B}}
\def\Bapp{\BB_\DD}

\def\C{\mathbb{C}}
\def\CC{{\cal C}}
\def\cun{{{\cal C}^1}}
\def\czero{{{\cal C}^0}}
\def\dklnp{\delta_{K,L}^{n+1}}
\def\Dhcarreuklnp{{{\dklnp (h^2(U))}}}
\def\DD{{\cal D}}

\def\EE{{\cal E}}
\def\TT{{\cal T}}
\def\UU{{\cal U}}

\def\ep{{\varepsilon}}

\def\R{\mathbb{R}}

\catcode`\@=11
\newif\ifproofmode
\proofmodetrue
\def\labelcour{nolabel}

\def\theequation{\thesection.\arabic{equation}}

\def\@endtheorem{\hfill\ifproofmode\rlap{\tiny \kern5mm
\labelcour}\fi\endtrivlist}

\catcode`\@=12 \proofmodefalse
\title{A mathematical model for the therapy of the HIV infection}

\author{Giulio Della Rocca\footnote{Math Department CSULB, 1250 Bellflower Blvd,
Long Beach CA 90840, USA, {\tt gdrocca@math.csulb.edu}} $\;$
Marco Sammartino\footnote{Corresponding author. Dipartimento di Matematica ed
Applicazioni, Universit\`a di Palermo,Via Archirafi 34, 90123 Palermo, ITALY,
{\tt marco@math.unipa.it}}$\;$
and Luciano Seta \footnote{ITD--CNR Palermo, Via Ugo La Malfa 153,
Palermo ITALY, {\tt seta@itd.cnr.it}}}

\date{}
\maketitle

\medskip

\begin{abstract}
In  \cite{SS} it was introduced a model to describe the dynamics
of the HIV infection when the patient is under chemotherapy
(either RTI or PI). The main idea in \cite{SS} was  to introduce
the effectiveness of the drug as a dynamical variable.

In this paper we pursue this idea starting from an analysis of the fitness of the virus during the
therapy.
We introduce an adaptive model in which the  ability of the viruses to infect the target cells is
related to the number of contacts between viruses and T--cells that have been inhibited by the drug.
This approach is similar to the model proposed in \cite{N82} for a predator-prey system.
However the biological interpretation is  different here because in our
context the adaptation of the virus is due
to the development of resistant virus strains.
We analyze different combination therapies with three antiviral drugs, which consist of
reverse transcriptase inhibitor (RTI) and protease inhibitor (PI) and
we show the possibility  of very long latency periods,  during which
the viral load goes below the detectable level.
These periods are followed by a rebound followed by the
re--establishing of the conditions previous to the therapy.
This dynamics is in good qualitative agreement with the
available clinical data.

%In a recent paper \cite{SS} was introduced a model to describe the
%dynamics of a HIV infection when the patient undergoes
%chemotherapy (either RTI or PI). The main idea is to introduce the
%effectiveness of the drug as a dynamical variable. In this paper
%we extend the analysis to a case of combination therapy, e.g.
%three-therapy. We show the possibility of very long latency
%periods.  During that period, the viral load goes below the
%detectable level. These periods are followed by a rebound, then
%followed by a re-establishing of the conditions previous to the
%therapy.
%This dynamics is in good qualitative agreement with the
%available clinical data.
\end{abstract}

\noindent
{\bf Math subject classification:} 92B05, 92C50, 34A99


\section{Introduction}
\setcounter{equation}{0}

The dynamics of the HIV infection is a complex process that
has many puzzling features.
The target of the HIV virus are the CD4+ T cells (referred as T-cells in this paper).
These cells, through the secretion of growth and differentiation
factors, play a crucial role in the immune response of the patient.
When the viral attack leads the level of the T-cells below a
critical level, the result is the immunodeficiency that characterizes
AIDS.
In the past 15 years several mathematical models have been formulated
to explain the development of the infection in dynamical terms
(see e.g. \cite{DH}, \cite{PKB}, \cite{EP}, \cite{K}, \cite{PN} and references
therein).
These models, together with the design of the appropriate experiments
and clinical trials, have significantly contributed to our
understanding of the processes underlying the HIV infection.
Reflecting on the fact that most HIV patients develop AIDS after several years with the infection, leads to the assumption that we are dealing with a slow dynamical time scale of the infection.
The mathematical modeling has significantly helped to clarify this assumption.
In fact, it is now widely recognized that the dynamics of the HIV
infection evolves through many different time scales.
Mathematical modeling helped in the uncovering of these time scales
and to the assessment that they are a result of
important biological processes involved in the infection.

In the mid 1990's, the development of potent anti-viral drugs began.
The combination of two, three and sometimes four of these drugs
resulted in highly active anti-retroviral therapies (HAART), that
changed the HIV infection from a fatal disease to an
illness that, in most cases, can be chronically managed for
several (up to 10-15 or more) years.
The introduction of HAART gave more momentum to mathematical modeling efforts
(see \cite{PN}, \cite{KW}, \cite{MBSN}, \cite{PECVHSMH} and
references therein) for several reasons.
In most patients under treatment, the viral load
decreased under the detectable level, leading to the hope that a carefully designed treatment could completely eradicate the infection.  Issues like:  {\em(a)}dose regimen control \cite{KLS},
{\em (b)} the appropriate choice of the starting time of the therapy,
{\em (c)} the appropriate choice of the combination therapy,
required a quantitative understanding of the dynamics of the
infection during the therapy.  We now know that complete virus eradication is still out of reach, thus increasing the role of mathematical modeling.  There is much debate in the medical community surrounding the failure of the therapies.
The existence of reservoirs (where the virus can survive and avoid
eradication, see e.g. \cite{HWP}), and the development of drug resistance
(see e.g. \cite{BN}) are the two mechanisms which have gained most credit in medical literature.

The perspective we shall use in this paper is based on the
assumption that the gradual mounting of virus resistance is the underlying
mechanism that drives the long term results of HAART.
To model the results of the virus resistance, some authors (see e.g. \cite{DBB}, \cite{BN}, \cite{SBD} and \cite{RB})
have theorized that two or more virus strains coexist in an infected patient:
one that is effected by the therapy; the wild-type, and  another that is resistant to therapy; the mutant strains.
Although these models have several merits, they encounter some difficulties.
The main problem is the fact that the introduction of new virus populations (the resistant strain)
necessitates the specification of a set of parameters seemingly difficult
to determine through clinical trials.

Our point of view will be different. We expect that the efficacy
of therapy evolves dynamically.
The ideas which our analysis is based
on, as expressed by \eqref{2.12} for the RTI therapy and
\eqref{2.13} for the PI therapy, are the following.
First,
following \cite{NBSM97}, we shall introduce the virus population
fitness and show how,  in presence of
the drug treatment, this fitness is related to the effectiveness of the drug.
Second, we shall assume that the fitness of
the virus population  depends on an
adaptive mechanism (due to the selective pressure introduced by
the presence of the drug).
We quantify the strength of the selective pressure as the total number
of the viruses inhibited by the drug action.
This is in analogy to \cite{N82} or \cite{N86}.
These assumptions will lead us to formulate our
model equations.

The paper is organized as follows:
In the next section we introduce the basic model introduced by Perelson
and his coworkers describing the dynamics of the infection without therapy.
After introducing the effect of the therapy as shown e.g. in \cite{PN}, we
derive an ODE for the effectiveness of the therapy.
This ODE has to be coupled with the equations ruling the dynamics of the $T$--cells and
of the viruses populations.
In Section 3, we show some numerical test where we simulate the effects
of the three-therapies (resulting from different
combinations of the RTIs and PIs),
In Section 4 we show that the parameters appearing in our model can be estimated
through a careful screening of the decline of the virus load during the
first weeks of treatment.
Finally we draw some conclusions.

\section{The drug effectiveness}

\subsection{The basic model}

The basic mechanism of the HIV infection is the following: a virus
penetrates a T-cell, uses the genetic material from the T-cell
to produce copies of itself, and eventually kill
the T-cell to release the newly produced viruses into the blood stream to infect other T-cells.

If one denotes with $T$ the plasma concentration of the T-cells,
with $T^*$ the concentration of the infected T-cells, and with $V$ the concentration of the HIV viruses,
one can write the following equations:
\begin{eqnarray}
\displaystyle\frac{dT}{dt}&=&\displaystyle s -\mu_T T+
rT\left(1-\frac{T+T^*}{T_{\textrm{\tiny max}}}\right)-kVT;\label{2.1}\\
\displaystyle\frac{dT^*}{dt}&=&\displaystyle kVT -\mu_{T^*}T^*;\label{2.2}\\
\displaystyle\frac{dV}{dt}&=&\displaystyle N\mu_{T^*}T^*-cV. \label{2.3}
\end{eqnarray}

In the above equations $s$ represents the rate of production of the T-cells.
For example, due to the bone marrow; $\mu_T$ is the death rate of the T-cells;
the logistic term in \eqref{2.1} describes the T-cells capability of self-reproduction; $k$ is the rate of infection of the T-cells per unit of virus;
$\mu_{T^*}$ is the rate at which the viruses kill the infected T-cells;
$N$ is the number of copies that each virus is able to produce using
the genetic material of the infected T-cell.

The above parameters show strong individual variability.
In the table below, we report a set of values that are most commonly
used in the literature and which we use throughout the rest of the
paper.
These values are therefore to be thought as mean values.

\begin{table}[!ht]
\caption{Parameters mean values.\label{tab:par}}
\begin{tabular}{lll}\hline
 s                        &\small Source term for CD4+T cells                        & \small 10 day$^{-1}$mm$^{-3}$ \\
$\mu_T$                   &\small Natural death rate of CD4+T cells                  & \small 0.02 day$^{-1}$ \\
 r                        &\small Growth rate of CD4+T cells                         & \small 0.03 day$^{-1}$ \\
$T_{\textrm{\tiny max}}$  &\small Maximal population level of CD4+T cells            & \small 1500 mm$^{3}$   \\
 k                        &\small Rate of the infection for CD4+T cells & \small 2.4$\times$10$^{-5}$ day$^{-1}$mm$^{3}$ \\
$\mu_{T^*}$               &\small Natural death rate of infected CD4+T cells         & \small 0.26 day$^{-1}$ \\
 N                        &\small Number of viruses produced by infected T--cells &\small 500 \\
 c                        &\small Natural death rate of viruses                      & \small 2.4 day$^{-1}$  \\\hline
\end{tabular} \\[0.5ex]
\end{table}

\begin{figure}
\begin{center}
\includegraphics[height=7.cm,width=13cm]{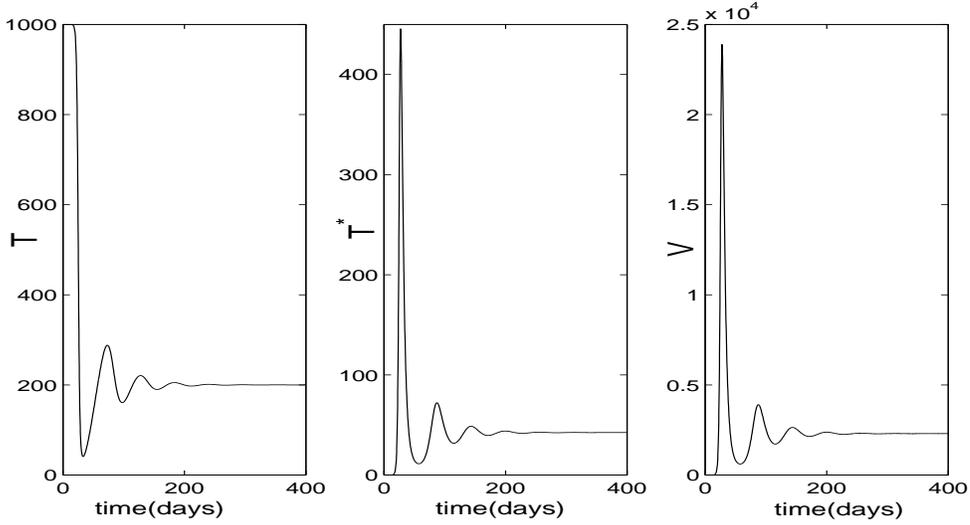}
\caption{The basic model without drugs: concentrations of
uninfected T-cells, infected T-cells and free viruses.}
\label{curu}
\end{center}
\end{figure}

In figure~\ref{curu} we have reported a simulation of the above
system, where we have taken the initial values
$T(0)=1000\, mm^{-3}$, $T^*(0)=0\, mm^{-3}$, $V(0)=10^{-3}$.

%%%%%%%%%%%% Figura 2 %%%%%%%%%%%%%%%%

\subsection{The therapies}

During the '90s, two anti-viral therapies were introduced.  One was based on Reverse Transcriptase Inhibitors (RTI).  These drugs are able to inhibit the Reverse Transcriptase enzyme.  This enzyme is necessary to the virus in order to use the DNA of the T-cell to replicate itself.  Therefore, when the enzyme is inhibited, a virus can penetrate a T-cell
but will not successfully infect it.
The model describing the dynamics of the infection under the influence of RTI therapy is:

\begin{eqnarray}
\displaystyle\frac{dT}{dt}&=&\displaystyle s -
\mu_T T+rT\left(1-\frac{T+T^*}{T_{\textrm{\tiny max}}}\right)-kVT; \label{rti1}\\
\displaystyle\frac{dT^*}{dt}&=&\displaystyle (1-\eta)kVT -\mu_{T^*}T^*; \label{rti2}\\
\displaystyle\frac{dV}{dt}&=&\displaystyle N\mu_{T^*}T^*-cV; \label{rti3}
\end{eqnarray}

In the above equation the parameter $\eta$ is the effectiveness of the drug.
When $\eta=1$ the drug is $ 100\%$ effective.

The second class of therapy is the Protease Inhibitors (PI).
When the process of protease is inhibited, the virions produced using the genetic
material of the T''cells  are unable to fully mature and thus unable to reproduce.
Eventually, these virions die out without contributing to the infection.
The model describing the action of the PI is:

\begin{eqnarray}
\displaystyle\frac{dT}{dt}&=&\displaystyle s -
\mu_T T+rT\left(1-\frac{T+T^*}{T_{\textrm{\tiny max}}}\right)-kVT; \label{pi1}\\
\displaystyle\frac{dT^*}{dt}&=&\displaystyle kVT -\mu_{T^*}T^*; \label{pi2} \\
\displaystyle\frac{dV}{dt}&=&\displaystyle (1-\delta)
N\mu_{T^*}T^*-cV.\label{pi3}
\end{eqnarray}
In this case the effectiveness of the drug is $\delta$.

\subsection{The drug effectiveness as a dynamical variable}

In the literature both $\eta$ and $\delta$ have been  considered as constant.
The process of virus resistance has been described through the
introduction of a new strain of viruses that are resistant to the drug.
This approach has several merits, but has the fundamental disadvantage of the
difficulty in estimating the parameters describing the new population.

Here we  follow a different approach. In \cite{NBSM97}, in absence of therapy,
the authors considered $kN \mu_{T^*}$ (what they called {\em the basic reproductive
ratio}) to be a direct measure of the fitness of the virus population.
The reason for this choice is apparent when one considers that the reproductive capability
of the viruses is proportional to both the capability to infect the $T$--cells (expressed
by the coefficient $k$) and to the number of viruses produced by each infected $T$--cell
(expressed by $N \mu_{T^*}$).

It is clear that when a therapy is introduced the fitness of the virus population
is now expressed by $(1-\eta)kN \mu_{T^*}$ and $(1-\delta)kN \mu_{T^*}$
for RTI and PI respectively.

We now suppose that the fitness of the viruses evolves dynamically in time
through an adaptive mechanism induced by the selective pressure.
In population dynamics (for example in the case of competing species),
to model adaptive mechanisms, one assumes that the fitness increases
with the total number (i.e. the integral with respect to time) of
the interaction between species. see e.g. \cite{N82}\cite{N86}.
By analogy we quantify the selective pressure induced by the drug as the total number
of viruses inhibited by the drug.
Taking $(1-\eta)$ and $(1-\delta)$ as measures of the fitness ($k$, $N$ and $\mu_{T^*}$
being constant), we write:

\begin{eqnarray}
    1-\eta(t)=ak\int\limits_0^t \eta(s)V(s)T(s)\ ds\, ; \label{2.12}\\
    1-\delta(t)=bN\mu_{T^*}\int\limits_0^t \delta(s)T^*(s)\ ds \, . \label{2.13}
\end{eqnarray}

The positive constants $a$ and $b$ represent the rate of growth of the
resistance for the specific drug, RTI and PI respectively.
In an adaptive population dynamics setting many authors, to model a loss
of memory effect, have introduced, inside the integral, a decaying kernel.
For sake of simplicity we don't consider this effect here.

The above formulas, that are integral equations, can be put in differential
form.
The resulting model equations for the drug effectiveness are respectively:
\begin{equation}
\displaystyle \frac{d\eta}{dt} =  -ak\eta VT \; , \label{eta}
\end{equation}
and
\begin{equation}
\displaystyle \frac{d\delta}{dt} =  -bN\mu_{T^*}\delta T^*\; . \label{delta}
\end{equation}

Equation \eqref{eta} has to be coupled with eqs.\eqref{rti1}--\eqref{rti3},
while eq.\eqref{delta} has to be coupled with eqs.\eqref{pi1}--\eqref{pi3}.

\section{The three--therapies}

At first, the results of the therapies were encouraging.
When a therapy starts, the infection appears to experience a period
of remission (that in the paper we shall call latency period),
during which the viral load decreases significantly, while the number
of the T-cells increases to the point where patients show no symptoms of the infection.
After this latency period, the viral load increases again and the condition
of the patient returns to the state before therapy.
The length of the latency period is variable (usually few weeks, for mono-therapy), but for bi-therapies
or three-therapies, these latency periods become several months
or several years long.
Even with multi-therapies, the latency period ends.
The hope that one could eradicate the infection with the multi-therapies
has been replaced for a more realistic goal of controlling the infection,
making it a chronic condition for as long as possible.

In this section we shall see how combination therapy of three drugs can
lead to a decline of the viral load and to the increase of T-cells
for periods of several thousands of days.
To the best of our knowledge, this is the first attempt of any modeling of the latency period and of the subsequent failure of the therapy.
We first consider the case of three-therapy based on the use of two RTIs
and one PI.
The system that models this association of drugs is:
\begin{eqnarray}
\displaystyle\frac{dT}{dt}&=&\displaystyle s -\mu_T T+
rT\left(1-\frac{T+T^*}{T_{\textrm{\tiny max}}}\right)-kVT;\\
\displaystyle\frac{dT^*}{dt}&=&\displaystyle (1-\eta_1)(1-\eta_2)kVT
-\mu_{T^*}T^*;\\
\displaystyle\frac{dV}{dt}&=&\displaystyle
(1-\delta)N\mu_{T^*}T^*-cV;\\
\displaystyle \frac{d\eta_1}{dt}& = & -a_1k\eta_1^3VT \; \\
\displaystyle \frac{d\eta_2}{dt}& = & -a_2k\eta_2^3VT \;  \\
\displaystyle \frac{d\delta}{dt}& = & -bN\mu_{T^*}\delta^3T^*\; .
\end{eqnarray}

In the above equations $\eta_1$ and $\eta_2$ are the effectivenesses of the two
Reverse Transcriptase Inhibitors, while $\delta$ is the effectiveness of the
Protease Inhibitor. The parameters that rule how fast the HIV virus develops
resistance to the drugs are $a_1$, $a_2$ and $b$ respectively.
In the simulation below we have taken $a_1=a_2=1.45\times10^{-3}$ and
$b=1.0\times10^{-5}$.

\begin{figure}
\begin{center}
\includegraphics[height=8cm,width=13cm]{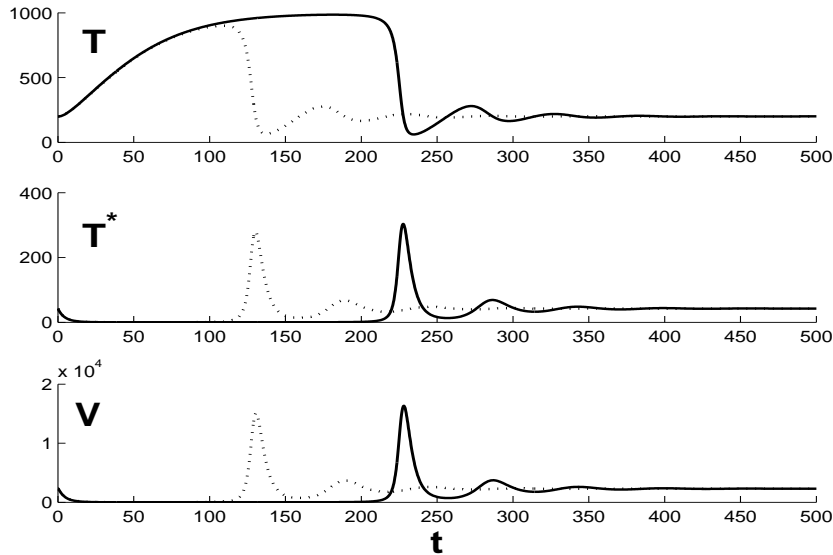}
\caption{
The dynamics of the infection with a bi-therapy with two RTI's.
The dashed line represents the case of the initial effectiveness of both drugs
is $\eta(t=0)=0.65$. The solid line represents the case
$\eta(t=0)=0.7$ for both drugs.}
\label{RTIRTI}
\end{center}
\end{figure}

For comparison purposes in Figure \ref{RTIRTI} we show the dynamics
of the infection when the treatment is bi-therapy with two RTIs.
The initial datum for $T$, $T^*$ and $V$ is  the steady state
of the system \eqref{2.1}--\eqref{2.3}.
This corresponds to starting the therapy when the infection is
in a mature state.
For the effectiveness we assumed that at the start of
therapy $\eta_1(t=0)=\eta_2(t=0)=0.65$ and $\eta_1(t=0)=\eta_2(t=0)=0.7$.
Through these simulations, one can appreciate how bi-therapy causes a rapid
decline of the virus load and an increase of healthy T-cells.
For a period of about three months (or about eight months when initially the drugs are more effective) the virus load decreases below the detectable level.
After this latency period, the virus load increases as the number of T-cells decline.  A rebound of the virus load appears until the count of both the virus and the T-cells return to the initial count before therapy.  All these findings from the model are consistent with the available clinical
data.

\begin{figure}
\begin{center}
\includegraphics[height=8cm,width=13cm]{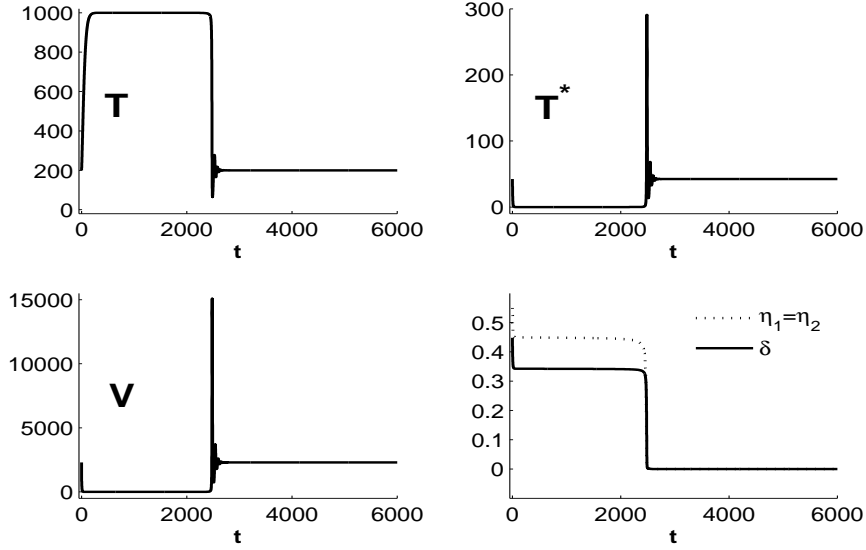}
\caption{
The dynamics of the infection with a three-therapy with two RTIs
and one PI.
The Figure represents the case when the initial effectiveness of both
the two RTIs is $\eta(t=0)=0.55$ while for the PI $\delta(t=0)=0.448$.
}
\label{RTIRTIPI}
\end{center}
\end{figure}

The situation is qualitatively the same for the case of the
three-therapy, but quantitatively different. In fact one can see
that adding a PI to the treatment, even with a low initial
effectiveness of $\delta(t=0)=0.448$, increases the
latency period dramatically. This is shown in Figure \ref{RTIRTIPI} where the behavior of the effectiveness of the three drugs is
represented. The latency period is about eight years long during
which the virus load is very low: $V\sim 0.1$ during the first
3-4 years of the therapy reaching the minimum a few weeks after starting the therapy.
Gradually, the virus load increases for several years, still remaining very low, until a
turning point is reached after which there is a violent rebound with wide oscillations.
Afterwards, the infection stabilizes at the pre-therapy level.
The effectiveness of the drugs, after an initial drop out, stabilizes for several years.
Perhaps, this is a result of the drastically low virus count after the initial stage of the therapy.
The selective pressure of the drugs on the virus population takes several years to induce significant
effects.
Finally, the effectiveness drops sharply and rapidly declines to zero.

We now examine the case of a three therapy resulting from
the association of one RTI and two PIs.
The system governing the dynamics is in this case:
\begin{eqnarray}
\displaystyle\frac{dT}{dt}&=&\displaystyle s -\mu_T T+
rT\left(1-\frac{T+T^*}{T_{\textrm{\tiny max}}}\right)-kVT;\\
\displaystyle\frac{dT^*}{dt}&=&\displaystyle (1-\eta)kVT
-\mu_{T^*}T^*;\\
\displaystyle\frac{dV}{dt}&=&\displaystyle
(1-\delta_1)(1-\delta_2)N\mu_{T^*}T^*-cV;\\
\displaystyle \frac{d\eta}{dt}& = & -ak\eta^3VT \; \\
\displaystyle \frac{d\delta_1}{dt}& = & -b_1N\mu_{T^*}\delta_1^3T^*\; .\\
\displaystyle \frac{d\delta_2}{dt}& = & -b_2N\mu_{T^*}\delta_2^3T^*\; .
\end{eqnarray}

\begin{figure}
\begin{center}
\includegraphics[height=8cm,width=13cm]{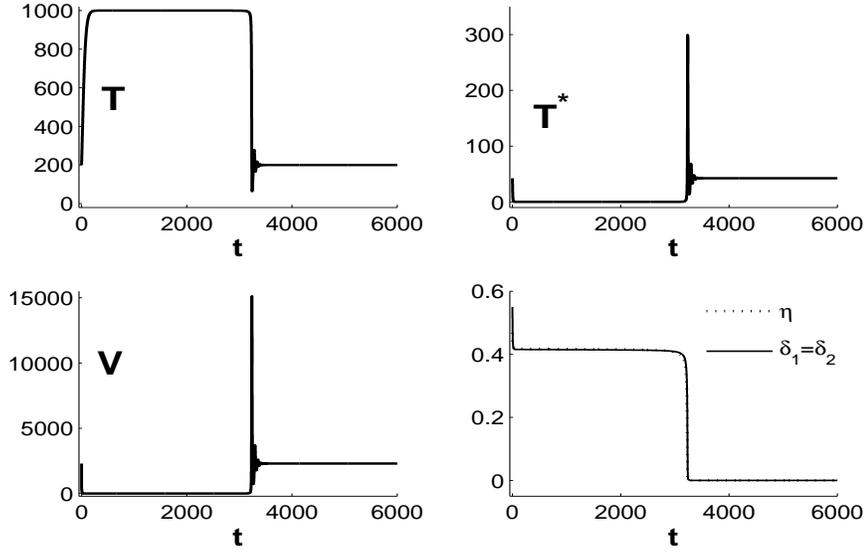}
\caption{
The dynamics of the infection with a three-therapy with two PIs
and one RTI.
The Figure represents the case when the initial effectiveness of
the two PIs is $\delta(t=0)=0.55$ while
for the RTI $\eta(t=0)=0.467$.}
\label{PIPIRTI}
\end{center}
\end{figure}

\begin{figure}
\begin{center}
\includegraphics[height=6cm,width=9cm]{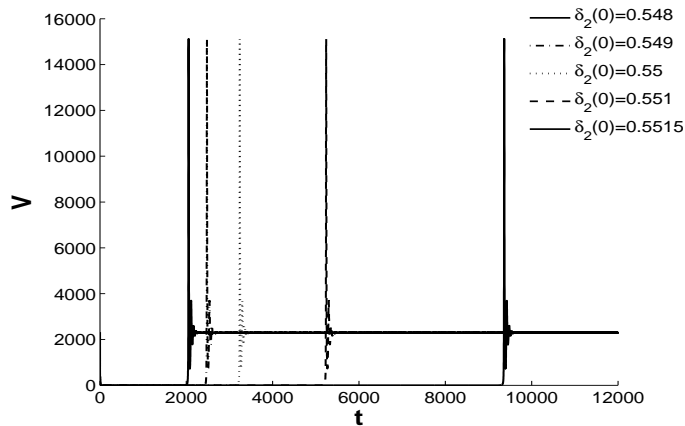}
\caption{
The virus load with different initial effectiveness.
In the Figure $\eta(t=0)=0.467$, $\delta_1(t=0)=0.55$ while
$\delta_2(t=0)=0.548, 0.549, 0.55, 0.551, 05515$.
The latency period shows a very sensitive dependence to the
initial effectiveness of the drugs.}
\label{latperiods}
\end{center}
\end{figure}

Some numerical results are shown in Figure \ref{PIPIRTI}.
In Figure \ref{latperiods} we represent the viral load
for different initial effectiveness of the drugs.
In the four simulations we keep $\eta(t=0)=0.467$, $\delta_1(t=0)=0.55$ while
we vary $\delta_2(t=0)$.
The latency period varies significantly with a slight change of the
initial effectiveness of one PI ranging from about 6 years to
about 28 years.

\section{Parameter estimation and the bi-phasic decline}

The main problem encountered when attempting to create a predictive model
from the equations discussed in the previous sections is the estimation of parameters
$a$ and $b$ (related to how fast the virus develops resistance to the
specific drug),  and the initial effectiveness of a specific drug.

\begin{figure}
\begin{center}
\includegraphics[height=6cm,width=8cm]{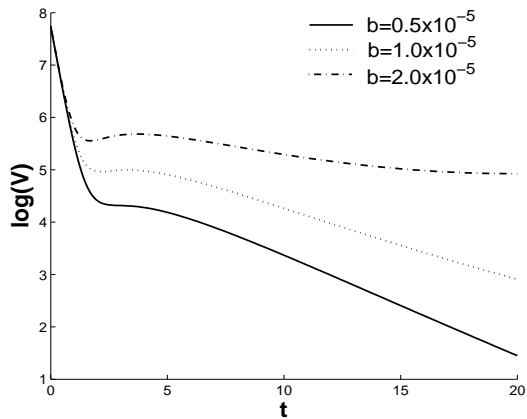}
\caption{
The behavior of the log of the virus load during the
first three weeks of a PI mono-therapy. A bi-phasic decline is
apparent. The beginning of the second phase of the decline depends on
the value of $b$.}
\label{diffbs}
\end{center}
\end{figure}

In Figure \ref{diffbs} one can observe the virus load (in logarithmic
scale) during the first three weeks of a treatment with one PI.
The three curves correspond to different $b$'s, the parameter ruling
how fast the virus develops resistance.
In all cases one observes a bi-phasic decline;
in both phases the
decline is exponential but with sharply different rates.
The bi-phasic decline has been observed during clinical trials and
our models can reproduce this feature.
In the picture, one can observe that the beginning of the second
(slower) phase, depends on $b$.

\begin{figure}
\begin{center}
\includegraphics[height=6cm,width=8cm]{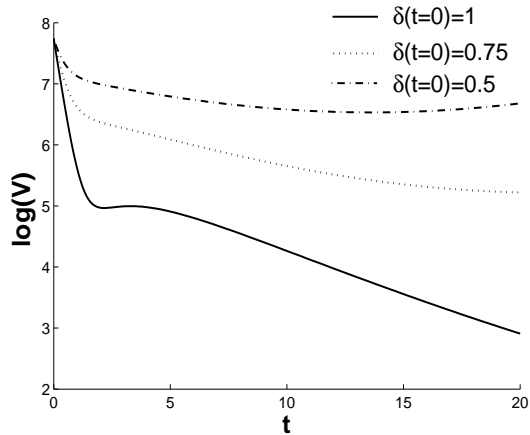}
\caption{
The behavior of the log of the virus load during the
first  week of a PI mono-therapy with different initial effectiveness.
The rate of the decline during the first phase depends on the
initial effectiveness. In the picture shown we have taken $b=1\times 10^{-5}$.}
\label{diffiniconds}
\end{center}
\end{figure}
In Figure \ref{diffiniconds} , again one can see the virus load for a
monotherapy with a PI, but with different initial effectiveness.
In this case one can appreciate that the rate of decline during the
first phase of the decline depends on the initial effectiveness.

One observes similar behaviors for the case of a therapy with a RTI.

The picture above suggests that a careful study of the behavior of the
virus load during the first weeks of a therapy could give an estimate
on the parameters ruling the dynamics of the infection, and therefore
an indication of the length of the latency period.

\section{Conclusions}

In this paper we have introduced a model to study the long term
behavior of HAARTs (highly active anti--retrovirus therapies).
The basic idea has been to
show how in presence of the treatment the fitness of the viruses is related
to the effectiveness of the drug.
Assuming that the fitness of the virus increases through an adaptive mechanism
we are able to derive an ODE (\eqref{eta} for the RTI and \eqref{delta} for
the PI) that rules the dynamics of the effectiveness of the drug.
The model, implemented in the case of the three-therapies, shows a good
qualitative agreement with the available clinical data.
During the first weeks of treatment the virus load experiences a rapid
bi-phasic decline, while the number of the T-cells reaches a level
ensuring good health conditions to the patient.
This phase is followed by a latency period during which the condition
of the patient remains stable for a period that can last for several years.
The length of the latency period depends on the initial effectiveness of the
drugs and on the parameters ruling how fast the virus develops resistance.
Eventually, the latency period ends and, following a violent virus rebound,
the infection returns to the condition preceding the therapy.
This behavior is the typical pattern reported in current medical literature.
We notice the features of this model are robust with respect to variations
in our model assumptions; e.g. one could assume a different power law for the
effectiveness in the right hand side of \eqref{eta} or \eqref{delta}
and get the same qualitative behavior of the dynamics, see \cite{SS}.

To the best of our knowledge this is the first model that describes the
whole course of the therapy, from the initial stages to the
insurgence of virus resistance leading to the inefficacy of the therapy.
Should this model be validated from a quantitative point of view,
it could be used as a tool to find the best strategies in modulating
the therapy to obtain longer latency periods whilst minimizing the side effects
of the drugs.
Apart from the validation of the model through clinical trials, many problems
remain. For example, one could ask what is the best moment to start the
therapy to make the latency period longer. Or take into account the fact
that the drug is taken by the patient discontinuously at specific hours
of the day (and not continuously as assumed by the model).
This and other topics will be the subject of future work.

\section*{Acknowledgments}

The work of the second (MS) and third (LS) authors has been supported
by the PRIN grant ``Nonlinear Mathematical Problems of
Wave Propagation and Stability in Models of Continuous Media''.


\begin{thebibliography}{38}

\bibitem{MBSN}
S.Bonhoeffer, M.May, G.M.Shaw and M.A.Nowak,
{\it Virus dynamics and drug therapy,}
Proc. Nat. Acad. Sci. USA, {\bf 94}, pp. 6971--6976 , (1997).

\bibitem{BN}
S.Bonhoeffer and M.A.Nowak,
{\it Pre--existence and emergence of drug resistance in HIV--1 infection,}
Proc. Roy. Soc. London B, {\bf 264}, pp. 631--637 , (1997).

\bibitem{DBB}
R.J.de Boer and C.A.B.Boucher,
{\it Anti--CD4 Therapy for AIDS Suggested by Mathematical Models,}
Proc. R. Soc. Lond., {\bf 263}, pp. 899--905, (1996).

\bibitem{DH}
J.Dolezal and T.Hraba,
{\em Application of mathematical model of immunological tolerance
to HIV infection,}
Folia Biol., {\bf 34}, pp. 336--341, (1988).

\bibitem{EP}
P.Essunger and A.S.Perelson,
{\it Modeling HIV infection of CD4+ T--cell subpopulations,}
J. Theoret. Biol., {\bf 170}, pp. 367--391, (1994).

\bibitem{HWP}
W.S.Hlavacek, C.Wofsy and A.S.Perelson,
{\it Dissociation of HIV--1 from follicular dendritic cells
during HAART: Mathematical analysis,}
PNAS, {\bf 96}(26), pp. 14681--14686, (1999).

\bibitem{K}
D.Kirschener,
{\it Using mathematics to understand HIV immune dynamics,}
Notices Amer. Math. Soc., {\bf 43}, pp. 775--792, (1996).

\bibitem{KLS}
D.Kirschener, S.Lenhart and S.Serbin,
{\it Optimal control of the chemotherapy of HIV,}
J. Math. Biol., {\bf 35}, pp. 775--792, (1997).

\bibitem{KW}
D.Kirschener and G.F.Webb,
{\it A model for treatment strategy in the chemotherapy of AIDS,}
Bull. Math. Biol., {\bf 59}, pp. 763--785, (1997).

\bibitem{N82}
V.W.Noonburg,
{\it Effects of behavioral adaptation on a predator--prey model,}
J. Math. Biol., {\bf 15}, pp. 239--247, (1982).

\bibitem{N86}
V.W.Noonburg,
{\it Competing species model with behavioral adaptation,}
J. Math. Biol., {\bf 24}, pp. 543--555, (1986).


\bibitem{NBSM97}
M.A.Nowak, S.Bonhoeffer, G.M.Shaw and R.M.May,
{\it Anti--viral drug treatment: Dynamics of resistance in free virus and infected
cell population,}
J. Theor. Biol.,{\bf 184}, pp. 203--217, (1997).

\bibitem{PECVHSMH}
A.S.Perelson, P.Essunger, Y.Cao, M.Vesanen, A.Hurley, K.Saksela,
M.Markowitz and D.D.Ho,
{\it Decay characteristics of HIV--1 infected compartments during
combination therapy,}
Nature, {\bf 387}, pp. 188--191, (1997).

\bibitem{PKB}
A.S.Perelson, D.Kirschener and R.De Boer,
{\it Dynamics of HIV infection of CD4+T cells,}
Math. Biosci., {\bf 114}, pp. 81--125, (1993).

\bibitem{PN}
A.S.Perelson and P.W.Nelson,
{\it Mathematical analysis of HIV--1 dynamics in vivo,}
SIAM Review, {\bf 41}, pp. 3--44 , (1999).


\bibitem{RB}
R.M.Ribeiro and S.Bonhoeffer,
{\it Production of resistant HIV mutants during antiretrovirial therapy,}
PNAS, {\bf 97}(14), pp. 7681--7686, (2000).

\bibitem{SS}
M.Sammartino and L.Seta,
{\it A model for the chemotherapy of the HIV infection with antigenic variation,}
Proceedings of WASCOM 2001, World Sci. Publishing, River Edge, NJ, pp. 521--526, (2002).

\bibitem{SBD}
N.I.Stilliakis, C.A.B.Boucher, M.D.De Jong, R. Van Leeuwen, R.Schuurman and R.J.De Boer,
{\it Clinical Data Sets of Human Immunodeficiency Virus Type 1 Reverse Transcriptase--Resistant Mutants Explained by a Mathematical Model,}
J. Virol., {\bf 71}(1), pp. 161--168, (1997).



\end{thebibliography}
\end{document}